# Broadband negative refraction of highly squeezed hyperbolic graphene plasmons

Jing Jiang,[1†] Xiao Lin,[1†*] Baile Zhang[1,2*]

[1]*Division of Physics and Applied Physics, School of Physical and Mathematical Sciences, Nanyang Technological University, Singapore 637371, Singapore.*

[2]*Centre for Disruptive Photonic Technologies, Nanyang Technological University, Singapore 637371, Singapore.*

[†]*These two authors contributed equally to this work.*

[*]*Corresponding authors. Email: xiaolinbnwj@ntu.edu.sg (X. Lin); blzhang@ntu.edu.sg (B. Zhang)*

**Negative refraction of highly squeezed polaritons is a fundamental building block for nanophotonics, since it can enable many unique applications, such as deep-subwavelength imaging. However, the phenomenon of all-angle negative refraction of highly squeezed polaritons, such as graphene plasmons with their wavelength squeezed by a factor over 100 compared to free-space photons, was reported to work only within a narrow bandwidth (<1 THz). Demonstrating this phenomenon within a broad frequency range remains a challenge that is highly sought after due to its importance for the manipulation of light at the extreme nanoscale. Here we show the broadband all-angle negative refraction of highly squeezed hyperbolic graphene plasmons in the infrared regime, by utilizing the nanostructured graphene metasurfaces. The working bandwidth can vary from several tens of THz to over a hundred of THz by tuning the chemical potential of graphene.**

# INTRODUCTION

Realizing negative refraction of highly squeezed polaritons, especially that supported by two-dimensional (2D) materials (*1-6*), such as graphene plasmon polaritons, is an important step toward the active manipulation of light at the extreme nanoscale (*7-9*), and can promise many photonic and optoelectronic applications (*10-17*). In 2017, the phenomenon of all-angle negative refraction between highly squeezed *isotropic* graphene plasmons and hexagonal boron nitride's (BN) phonon polaritons, with their in-plane polaritonic wavelengths squeezed by a factor over 100, is theoretically shown possible in the graphene-BN heterostructures (*18*). By following Ref. (*18*), henceforth we use the parameter of squeezing factor to define the ratio between the photon wavelength in free space and the in-plane polaritonic wavelength. The extreme spatial confinement of these highly squeezed isotropic polaritons, however, also limits our ability to tailor their dispersion relations (18, *19*); in other words, their effective negative refractive index exists only within a certain frequency range. Consequently, the phenomenon of all-angle negative refraction of these highly squeezed isotropic polaritons is restricted to only a narrow frequency range (near 23 THz) with a bandwidth of less than 1 THz (*18*). The narrow bandwidth unavoidably hinders the demonstration of this exotic phenomenon in experiments and thus the potential applications. It remains a challenge to realize the all-angle negative refraction of highly squeezed polaritons within a broad frequency range.

Here we predict the phenomenon of broadband all-angle negative refraction of highly squeezed *hyperbolic* polaritons, with their squeezing factors over 100, in the infrared regime. The highly squeezed hyperbolic polaritons can be supported by nanostructured 2D materials such as graphene nanoribbon arrays (*20, 21*) or BN nanoribbon arrays (*22, 23*), by 2D materials supported by nanostructured substrates (*13, 24*), or by naturally existing hyperbolic 2D materials such as black phosphorous (*2, 25*). The peculiar isofrequency contour of hyperbolic polaritons (*2, 20, 26*) gives us an extra freedom to tailor their in-plane propagation direction, and thus the flexibility to realize the all-angle negative refraction of highly squeezed polaritons in a wide frequency range. As an example, for the all-angle negative refraction of hyperbolic graphene plasmons shown below, the working bandwidth can vary from several tens of THz to over a hundred of THz by simply tuning the chemical potential of graphene. Our work thus indicates that 2D materials are a versatile platform for the design of advanced nano-metasurfaces and nano-imaging elements, which are of fundamental importance for the active control of light at the nanoscale.

We note that there are other types of negative refraction studied in the platform of 2D materials, including the negative refraction of electrons (*27-29*), negative refraction of light propagating through a monolayer graphene (*30*), plasmonic (*31, 32*) and optical (*33, 34*) negative refraction in 3D bulk materials (i.e., 2D material-based periodic structures). However, these electromagnetic refractions [30-34] occur out of the plane of 2D materials and not in plane, thus without taking advantage of the high spatial confinement of polaritons that is critical in this work. In addition, the all-angle negative refraction of graphene plasmons was studied in the hybrid graphene-photonic crystal structures (*35*), but the working bandwidth is still far less than 1 THz for a given chemical potential of graphene, along with the squeezing factor less than 20.

**RESULTS AND DISCUSSION**

To highlight the underlying physics, we begin with the dispersion of hyperbolic polaritons supported by a uniaxial metasurface. The metasurface, such as that in the left region of Fig. 1A, can be modelled by an effective anisotropic surface conductivity of $\bar{\bar{\sigma}}_l = [\sigma_{xx,l}, \sigma_{yy,l}]$. Contrary to the isotropic metasurface (i.e., $\sigma_{xx,l} = \sigma_{yy,l}$) which supports the propagation of either transverse-magnetic (TM) (*36, 37*) or transverse-electric (TE) polaritons (*38-41*), the uniaxial metasurface supports the hybrid TM-TE polaritons (*20, 24, 42-44*). From the electromagnetic theory (*45*), the dispersion relation for hybrid polaritons (*42-44*) is derived as

$$(1 + \frac{k_{z1}/\varepsilon_{r1}}{k_{z2}/\varepsilon_{r2}}) + (\sigma_{xx,l}sin^2\phi + \sigma_{yy,l}cos^2\phi)\frac{k_{z1}/\varepsilon_{r1}}{\omega\varepsilon_0} = \frac{(\sigma_{xx,l}-\sigma_{yy,l})sin^2\phi cos^2\phi \frac{k_{z1}/\varepsilon_{r1}}{\omega\varepsilon_0}}{(\sigma_{xx,l}cos^2\phi +\sigma_{yy,l}sin^2\phi)+\frac{k_{z1}+k_{z2}}{\omega\mu_0}} \quad (1)$$

In the above, $\varepsilon_{r1,2}$ is the relative permittivity of region above/below the metasurface; $k_{z1,2} = \sqrt{\frac{\omega^2}{c^2}\varepsilon_{r1,2} - k_x^2 - k_y^2}$ and $\bar{k}_{||} = \hat{x}k_x + \hat{y}k_y$ are the out-of-plane and in-plane wavevectors of the hybrid polariton, respectively; $\phi$ is the angle between $\bar{k}_{||}$ and $\hat{x}$; $\omega$ is the angular frequency; $\varepsilon_0$, $\mu_0$ and $c$ are the permittivity, permeability, and light speed in free space, respectively. It is worthy to note that the left side of equation (1) is only relevant to the TM field components of hybrid polaritons, while the denominator of the right side of equation (1) only corresponds to the TE field components. To some extent, the numerator of the right side of equation (1) denotes the coupling strength between TM and TE field components.

Importantly, for the highly squeezed hybrid polaritons (i.e., $|\bar{k}_{||}| \gg \omega/c$), the right side of equation (1) is shown to be much smaller than 1, due to the difference of the spatial confinement between pure TE and TM polaritons; see Supplementary Materials. This way, the dispersion relation of hybrid polaritons in equation (1) can be compactly reduced to (*20, 24*)

$$(1 + \frac{k_{z1}/\varepsilon_{r1}}{k_{z2}/\varepsilon_{r2}}) + (\sigma_{xx,l}\sin^2\phi + \sigma_{yy,l}\cos^2\phi)\frac{k_{z1}/\varepsilon_{r1}}{\omega\varepsilon_0} = 0 \qquad (2)$$

In other words, the highly squeezed hybrid TM-TE polaritons are dominant by the TM field components. The isofrequency contour governed by equation (2) is hyperbolic, when $Im(\sigma_{xx,l}) \cdot Im(\sigma_{yy,l}) < 0$.

Figure 1B shows the hyperbolic isofrequency contour of highly squeezed polaritons supported by a graphene metasurface at 15 THz. As a conceptual demonstration, here we use the nanostructured graphene, i.e., graphene nanoribbon arrays, to create the hyperbolic metasurface; see the schematic structure in Fig. 1A. According to Ref. (*20*), when the pitch $L$ of nanoribbons is much smaller than the polaritonic wavelength $\lambda_{\text{polariton}}$, i.e., $L \ll \lambda_{\text{polariton}}$, the effective medium theory can be applied to describe the graphene metasurface. Then the effective anisotropic surface conductivity of graphene metasurface can be described by $\sigma_{xx,l} = \frac{L\sigma_s\sigma_C}{W\sigma_C + (L-W)\sigma_s}$ and $\sigma_{yy,l} = \sigma_s\frac{W}{L}$, where $W$ is the width of nanoribbon, $\sigma_C = -i(\omega\varepsilon_0 L/\pi)\ln[\csc(\pi(L-W)/2L)]$ is an equivalent conductivity associated with the near-field coupling between adjacent nanoribbons, and $\sigma_s$ is the surface conductivity of monolayer graphene modelled by the Kubo formula (*20, 46*). Here the nanostructured graphene has a chemical potential of 0.1 eV, a conservative electron mobility of 10000 cm$^2$V$^{-1}$s$^{-1}$ (*36, 37*), a pitch of $L = 30$ nm, and a width of $W = 20$ nm. The region below/above the graphene metasurface is the dielectric substrate (e.g., SiO$_2$) and air, respectively.

For the emergence of refraction phenomenon, the graphene metasurface in the right region in Fig. 1A shall be different from the left region. One simple way is to rotate the graphene metasurface in the right region with a certain angle with respect to the left region, such as 90$^0$ shown in Fig. 1A. This way, the graphene metasurface in the right region can be characterized by a surface conductivity of $\bar{\bar{\sigma}}_r = [\sigma_{xx,r}, \sigma_{yy,r}]$, where $\sigma_{xx,r} = \sigma_{yy,l}$ and $\sigma_{yy,r} = \sigma_{xx,l}$. By applying the conservation law for wavevectors parallel to the boundary between left and right regions (i.e., parallel to $\hat{y}$), the y-component of group velocity for the incident hyperbolic polaritons in the left region is opposite to that for the transmitted hyperbolic polaritons in the right region, as shown in Fig. 1B. This enables the negative refraction between hyperbolic polaritons in the left and right regions. In addition, we note that the hyperbolic polaritons in the left region can only propagate within a certain range of directions. Figure 1B indicates that for these hyperbolic polaritons (i.e., incident from the left region with arbitrary angles), negative refraction can always happen at the boundary. Here we denote this phenomenon as the all-angle negative refraction [18, 47, 48] of hyperbolic polaritons (i.e., graphene plasmons).

Figure 1C numerically demonstrates the all-angle negative refraction of hyperbolic polaritons at 15 THz, by using the finite-element method (COMSOL Multiphysics). Hyperbolic polaritons are excited by a dipole source in the left region and propagate directionally towards the boundary. At the boundary, the polaritonic beams are negatively refracted. Moreover, these beams converge to form an image in the right region, which computationally validates the all-angle negative refraction.

Here the squeezing factor $\frac{Re(|\bar{k}_{||}|)}{\omega/c} > 100$ [see Fig. 1B] indicates that when compared with the wavelength in free space, the polaritonic wavelength is squeezed at least by a factor over 100. It is worthy to emphasize that the all-angle negative refraction of highly squeezed hyperbolic polaritons with directional propagation, as an important advantage over that of isotropic polaritons which propagate omnidirectionally (*18, 47, 48*), might facilitate the design of novel compact guidance. We note that the imaging mechanism here is not a perfect image recovery. This is because the reflection at the boundary is unavoidable due to impedance mismatch between the two hyperbolic metasurfaces, and the propagation loss of hyperbolic polaritons will degrade the imaging quality; see Supplementary Materials. To optimize the quality of formed image in the right region, one shall consider both reflection and the propagation loss of polaritons (*18, 47, 48*).

Figure 2 shows the effective anisotropic conductivity of graphene metasurface, which can help to infer the working frequency range of the all-angle negative refraction of hyperbolic polaritons. As shown in Fig. 2A, the frequency range, that has $Im(\sigma_{xx,l}) \cdot Im(\sigma_{yy,l}) < 0$ and thus supports the hyperbolic polaritons, spans from 0 to 48 THz, when the chemical potential of graphene is 0.1 eV. It shall be noted that the phenomenon of all-angle negative refraction of hyperbolic polaritons can happen at arbitrary frequency within this frequency range, as long as the effective medium theory for graphene metasurface is valid (i.e., when $L \ll \lambda_{\text{polariton}}$). To guarantee the validity of the effective medium theory for metasurfaces based on 2D materials, a small value of pitch $L$, which although might increase the complexity in structural fabrication (*22, 23*), can be adopted. This way, the working frequency range for the all-angle negative refraction of highly squeezed polaritons revealed here is no longer limited by the frequency range supporting negative-index polaritons (such as Ref. (*18*)). Consequently, the working frequency range can be broadband and actively controllable via tuning the chemical potential of 2D materials.

Figure 2B shows the working bandwidth of all-angle negative refraction in graphene metasurfaces. The bandwidth of graphene metasurface having $Im(\sigma_{xx,l}) \cdot Im(\sigma_{yy,l}) < 0$ changes from 40 THz to over 100 THz, by increasing the chemical potential from 0.1 eV to 0.5 eV. Therefore, in principle, the bandwidth of all-angle negative refraction of hyperbolic polaritons can vary from several tens of THz to even over a hundred THz, by simply increasing the chemical potential of graphene. Interestingly, Fig. 2B shows that the loss can increase the bandwidth when $\mu_c$ is smaller than 0.18 eV; see analysis in Supplementary Materials.

To numerically validate the all-angle negative refraction of hyperbolic polaritons in a broad bandwidth, Fig. 3 demonstrates this phenomenon at other frequencies, i.e., at 10 THz in Fig. 3A and at 20 THz in Fig. 3B. The squeezing factors for hyperbolic polaritons at these two frequencies are both over 100; see Supplementary Materials. Therefore, by using hyperbolic metasurfaces based on 2D materials, we can extend the working bandwidth of all-angle negative refraction of highly squeezed polaritons to at least several tens of THz, which is favored for practical applications.

To further extend the bandwidth of all-angle negative refraction of highly squeezed polaritons, one may adopt the naturally anisotropic 2D materials to support tunable hyperbolic polaritons, such as those described in Ref. (*2*) including black phosphorous. From Ref. (*2*), these 2D materials can be directly characterized by an anisotropic surface conductivity, i.e.,

$$\sigma_{jj} = \frac{ie^2}{\omega + i/\tau} \cdot \frac{n}{m_j} + s_j \left[\Theta(\omega - \omega_j) + \frac{i}{\pi}\right], j = x, y \qquad (3)$$

where $n$ is the concentration of electrons, $m_j$ is the electron's effective mass along the $j$ direction, $\tau$ is the relaxation time, $\omega_j$ is the frequency of the onset of interband transitions for the $j$ component of conductivity, $s_j$ accounts for the strength of interband component, and $\Theta(\omega - \omega_j)$ is a step function. Equation (3) circumvents the requirement (i.e., $L \ll \lambda_{\text{polariton}}$ discussed in the above) for the validity of effective medium theory for metasurfaces based on nanostructured 2D materials. For these anisotropic 2D materials, the value of $n$ can be flexibly tunable via electrostatic gating, just like graphene, and $\omega_x$ can be different from $\omega_y$. These, along with the broad class of anisotropic materials, give us the flexibility to realize the all-angle negative refraction of highly squeezed polaritons in a broad frequency range.

In conclusion, we have revealed a viable way to realize the all-angle negative refraction of highly squeezed polariton in a broadband infrared regime, by utilizing hyperbolic metasurfaces based on 2D materials. Due to the combined advantages of highly directional propagation, active tunability, low loss and ultrahigh confinement provided by hyperbolic polaritons in 2D materials, the broad class of 2D materials can provide a versatile platform for the manipulation of light-matter interaction at the extreme nanoscale and for the design of highly compact nano-devices and circuits.

**MATERIALS AND METHODS**
The finite element simulation is implemented via the frequency domain simulation in the commercial software of COMSOL Multiphysics. To enable the high calculation accuracy, graphene is modeled as a surface, where an anisotropic surface conductivity is used to fulfill the conditions for discontinuities in the electromagnetic fields. The meshing resolution in the plane of graphene metasurface is 5 nm. For Figs. 1C&3, a z-polarized dipole source is placed in the left region at 5 nm above the graphene metasurface; the phenomenon of all-angle negative refraction is pronounced, independent of the vertical position of the source. The fields in Figs. 1C&3 are obtained at the plane with 2 nm above the graphene metasurface. For the clarity of conceptual demonstration, the value of $Re(\sigma_{yy,l})$ is artificially set to be equal to $Re(\sigma_{xx,l})$ in Figs. 1C&3. For the cases considering the realistic material loss, the phenomenon of all-angle negative refraction is shown in Fig. S2. The phenomenon of all-angle negative refraction is also independent of the relative permittivity of the dielectric substrate $\varepsilon_{r2}$, as shown in Fig. S4; in the main text, we set $\varepsilon_{r2} = 3.6$.

**SUPPLEMENTARY MATERIALS**
Section S1: Dispersion of hybrid polaritons supported by anisotropic metasurfaces
Section S2: All-angle negative refraction of hyperbolic graphene plasmons
Section S3: Loss influence on the bandwidth having $Im(\sigma_{xx}) \cdot Im(\sigma_{yy}) < 0$
Fig. S1. Real part of effective surface conductivity of graphene metasurface.
Fig. S2. All-angle negative refraction of hyperbolic polaritons when the real material loss is considered.
Fig. S3. Isofrequency contours of hyperbolic graphene plasmons at 10 THz, 15 THz and 20 THz.
Fig. S4. Substrate influence on the all-angle negative refraction of hyperbolic graphene plasmons at 15 THz.
Fig. S5. Loss influence on the bandwidth having $Im(\sigma_{xx,l}) \cdot Im(\sigma_{yy,l}) < 0$.
Fig. S6. Tunability of the negative refraction between hyperbolic graphene plasmons at 15 THz.

**REFERENCES AND NOTES**
1. D. N. Basov, M. M. Fogler, F. J. García de Abajo, Polaritons in van der Waals materials. *Science*

**Acknowledgements:** We thank Y. Yang, I. Kaminer and M. Soljačić for useful discussions. **Author contributions:** X.L. conceived the research. J.J. performed the main calculation. X.L. and B.Z. contributed insight and discussion on the results. J.J., X.L. and B.Z. wrote the paper. X.L. and B.Z. supervised the project. **Competing interests:** The authors declare that they have no competing interests. **Data and materials availability:** All data needed to evaluate the conclusions in the paper are present in the paper and/or the Supplementary Materials. Additional data related to this paper may be requested from the authors.

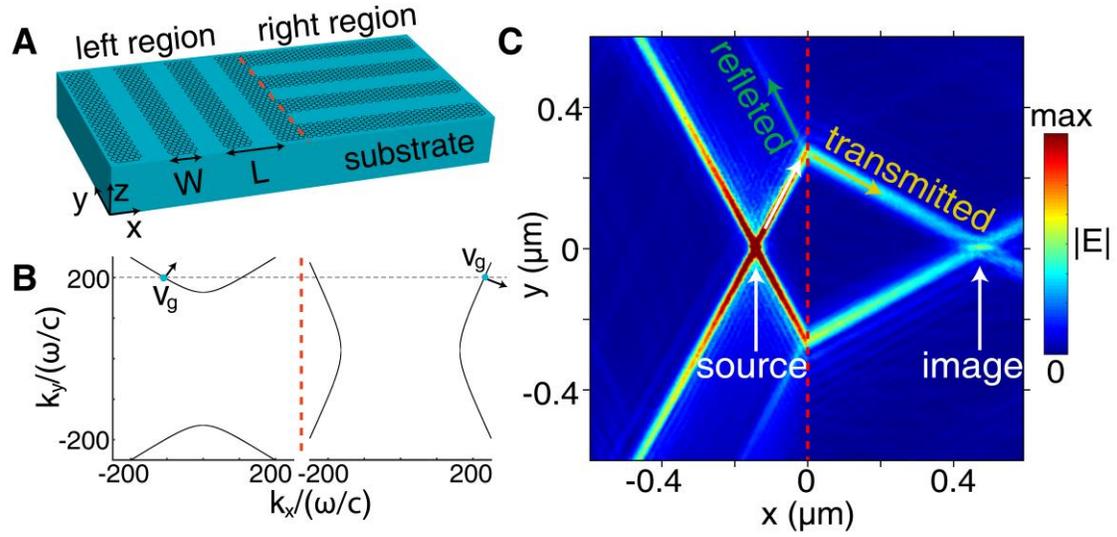

**Fig. 1. All-angle negative refraction of highly squeezed polaritons supported by hyperbolic metasurfaces.** (**A**) Structural schematic. The hyperbolic metasurfaces can be created by anisotropic 2D materials (such as black phosphorous) or nanostructured 2D materials (such as graphene nanoribbon array here). (**B**) Isofrequency contours of hyperbolic graphene plasmons, supported by metasurfaces in left/right regions in (A). The dashed grey line represents the condition for conservation of wave vectors parallel to the boundary. The arrows represent the directions of group velocity, indicating that for polaritons in the left region incident with arbitrary angle, negative refraction can happen at the boundary. (**C**) Distribution of electric field $|\bar{E}|$ excited by a dipole source. The red dashed line represents the boundary between left/right regions. Here, and in the figures below, the nanostructured graphene has a conservative electron mobility of 10000 $cm^2V^{-1}s^{-1}$, a pitch of $L = 30$ nm, and a width of $W = 20$ nm. The chemical potential of graphene is 0.1 eV. The working frequency is 15 THz.

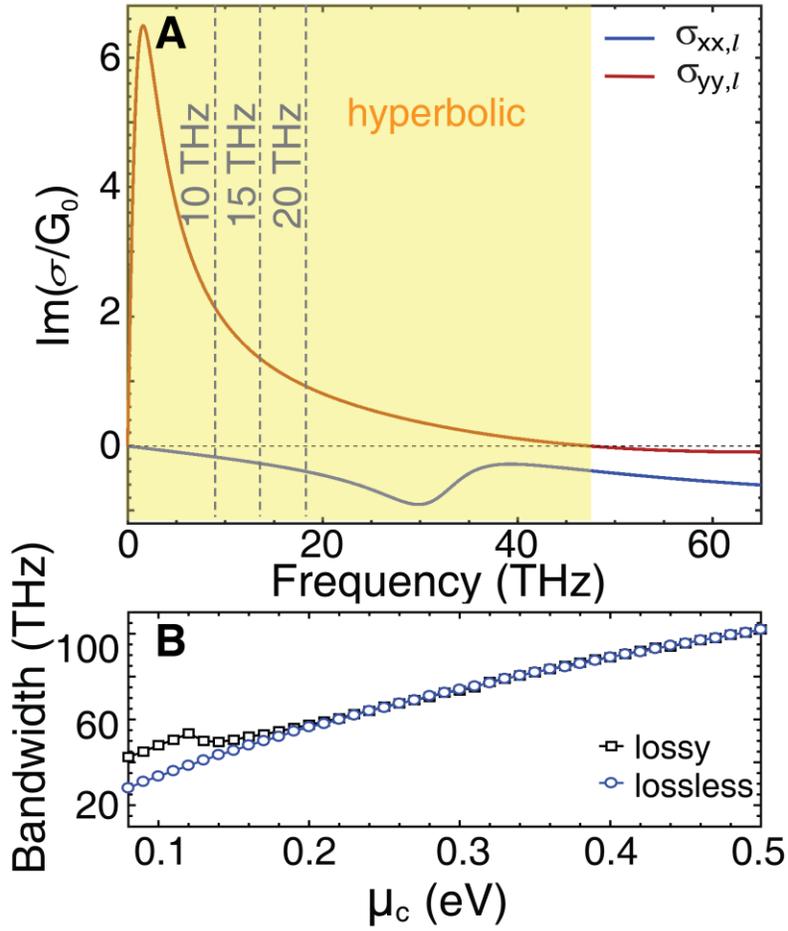

**Fig. 2. Broadband all-angle negative refraction of hyperbolic polaritons.** (**A**) Imaginary part of surface conductivity of graphene metasurface, created by nanostructured graphene as shown in the left region of Fig. 1A and having an effective anisotropic surface conductivity of $\bar{\bar{\sigma}}_l = [\sigma_{xx,l}, \sigma_{yy,l}]$. The graphene metasurface supports hyperbolic plasmon polaritons when $Im(\sigma_{xx,l}) \cdot Im(\sigma_{yy,l}) < 0$, i.e., the region highlighted by yellow. The chemical potential of graphene is $\mu_c = 0.1$ eV. (**B**) Bandwidth of graphene metasurface having $Im(\sigma_{xx,l}) \cdot Im(\sigma_{yy,l}) < 0$, as a function of the chemical potential. The phenomenon of all-angle negative refraction of hyperbolic polaritons can happen within this bandwidth. The constant $G_0 = e^2/4\hbar$ is the universal optical surface conductivity.

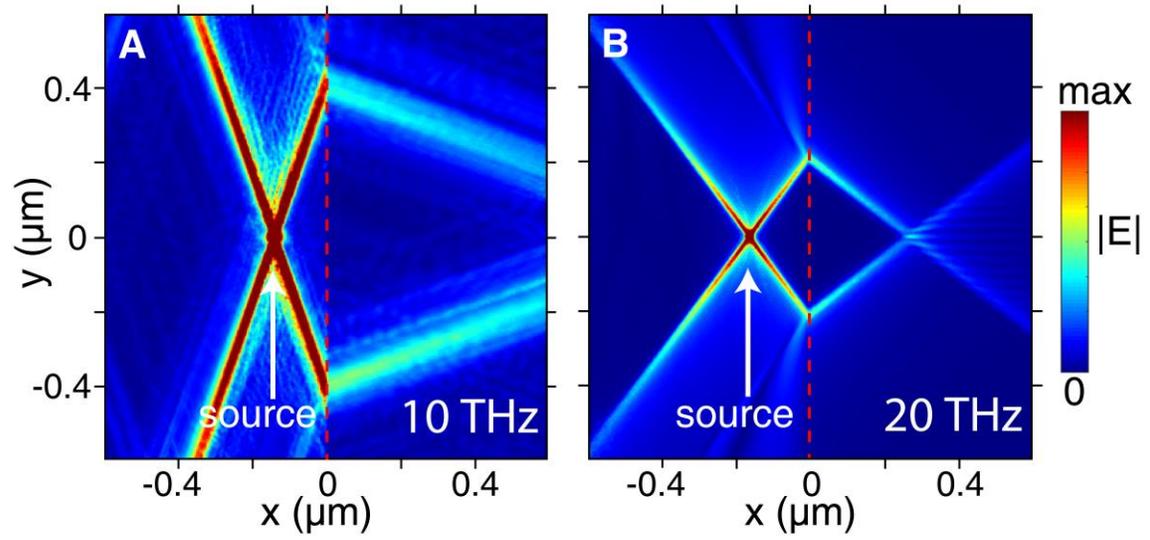

**Fig. 3. All-angle negative refraction of hyperbolic polaritons (A) at 10 THz and (B) at 20 THz.** The other parameters are the same as that in Fig. 1C. The value of anisotropic surface conductivity for graphene metasurfaces are highlighted by grey dashed lines in Fig. 2A.